\documentclass[aps,prl,twocolumn,showpacs,superscriptaddress,floatfix,preprintnumbers,amsmath,amssymb]{revtex4}

\usepackage{graphicx}
\usepackage{dcolumn}
\usepackage{bm}

\begin{document}

\preprint{Ver. 4.1}

\title{Tailoring exchange interactions in engineered nanostructures: \textit{Ab initio study}}

\author{O. O. Brovko}
\author{P. A. Ignatiev}
\author{V. S. Stepanyuk}
\affiliation{Max-Planck-Institut f\"{u}r Mikrostrukturphysik, Weinberg 2, D06120 Halle, Germany}
\author{P. Bruno}
\affiliation{Max-Planck-Institut f\"{u}r Mikrostrukturphysik, Weinberg 2, D06120 Halle, Germany}
\affiliation{European Synchrotron Radiation Facility, BP 220, F38043 Grenoble Cedex, France}

\date{\today}

\begin{abstract}
    We present a novel approach to spin manipulation in atomic-scale 
nanostructures. Our \textit{ab initio} calculations clearly demonstrate that it is 
possible to tune magnetic properties of sub-nanometer structures by adjusting the 
geometry of the system. By the example of two surface-based systems we 
demonstrate that (i) the magnetic moment of a single adatom coupled to a buried 
magnetic Co layer can be stabilized in either a ferromagnetic or an 
antiferromagnetic configuration depending on the spacer thickness. It is found 
that a buried Co layer has a profound effect on the exchange interaction between 
two magnetic impurities on the surface. (ii) The exchange interaction between 
magnetic adatoms can be manipulated by introducing artificial 
nonmagnetic Cu chains to link them.
\end{abstract}


\maketitle

    The interest in spin-based electronic devices of atomic dimensions has grown 
immensely as science and technology approached the atomic limit. In such systems, that are usually referred to as spintronic ones, a key role is played by magnetic interactions between single sub-nanoscale units. This makes gaining control over magnetic ordering down to the atomic level a task of major importance \cite{QuantComp}.

    The technological breakthrough that brought about gigabyte magnetic memory 
devices was enabled by the discovery of the GMR effect by P.~Gr\"unberg and 
A.~Fert \cite{Grunberg2,Fert}. Theoretical investigations have revealed that GMR 
has a common source with another interesting effect, the effect of an oscillatory 
interlayer exchange coupling 
(IEC)\cite{Grunberg2,PhysRevB.53.R2956,BrunoIEC_PRL91,BrunoIEC_PRB92}. Both of them originate in the spin-selective scattering of conductance electrons at magnetic layers. If two magnetic layers are separated with a non-magnetic metallic spacer, the conduction electrons scattered at each magnetic layer are forced to interfere and form standing waves inside the spacer. Since the scattering of minority and majority electrons at magnetic layers is different, the distributions of electronic densities in systems with parallel and antiparallel alignments of the spins also should be different. As a result, the system's ground state energy and spin configuration strongly vary with the thickness of the spacer, resulting in the oscillations of the interlayer exchange coupling \cite{BrunoIEC_PRL91,BrunoIEC_PRB92}. This reasoning is not restricted to the case of monolayers but can be applied to any kind of magnetic impurities coupled via conductance electrons. Affecting the interference of conduction electrons scattered at each impurity, e.g. by introducing structural changes in the metallic host,  one can tune the exchange interaction, similarly to the IEC. Nowadays modern experimental techniques such as the scanning tunneling microscopy (STM) make it possible to build in an atom-by-atom fashion complex surface nanostructures with predefined positions of magnetic impurities \cite{Stroscio06,Folsch07}. P
robing of the exchange interaction was only recently enabled by state of the art spin-flip \cite{Hirjibehedin:Nature312}, Kondo measurement \cite{Kern07} and magnetization curve mapping \cite{FockoMeier04042008} experiments, thus opening a new playground for studies of magnetic phenomena at the atomic scale.

    In this Letter we present a novel approach to controlling single spins in nanostructures adsorbed on metallic surfaces. We demonstrate that exchange coupling of adatoms and addimers to a magnetic layer across a nonmagnetic spacer displays an oscillatory behavior. This allows one, by deliberate choice of the spacer's thickness, to control the magnetic configuration and exchange interaction of single magnetic adatoms, driving them into either a ferro- or an antiferromagnetic behavior or even suppressing their magnetic properties. We also demonstrate that the quantum interference of electrons in quasi-one-dimensional engineered nanostructures, such as the linear atomic chains on a surface \cite{Stroscio06,Folsch07}, can be used to the same end as the quantum confinement in an overlayer. According to our calculations, by linking magnetic adatoms with nonmagnetic chains the exchange coupling between them can be significantly enhanced in magnitude, quenched or even reversed in sign at separations reaching up to $20~\mathrm{\AA}$.

    In our calculations we use the Korringa-Kohn-Rostoker (KKR) Green's function method in atomic spheres approximation  \cite{Stepanyuk.PRL.1995,Zeller.PRB.1995}. This method is based on the density functional theory in local spin density approximation. The KKR approach exploits the properties of the Green's function of the Kohn-Sham operator, in particular the fact that the electronic density can be expressed through the imaginary part of the energy-dependent Green's function of the system. An arbitrary system can be regarded as the perturbation of an ideal one with a known Green's function: the Green's functions of those two systems can be linked through the Dyson equation \cite{zab05}. We treat a surface as a 2D perturbation of an ideal crystal bulk with a slab of vacuum. Taking into account the translational symmetry of the surface geometry, the Green's functions are formulated in momentum space. Adatoms and chains are considered as the perturbation of the clean surface. These calculations are performed in real space. Exchange energies of compact systems (where the charge redistribution is significant, $d\lesssim5~\mathrm{\AA}$) are calculated as the difference between the total energies of fully self-consistent ferromagnetic and antiferromagnetic configurations. For large distances between constituent system's parts the interaction energies are calculated according to the force theorem \cite{Lang96,Stepanyuk04} from the single-particle energies alone \cite{Hylgaard00}. This approach allows to resolve small energy differences with high accuracy \cite{Stepanyuk05}.

    \begin{figure}
		\includegraphics{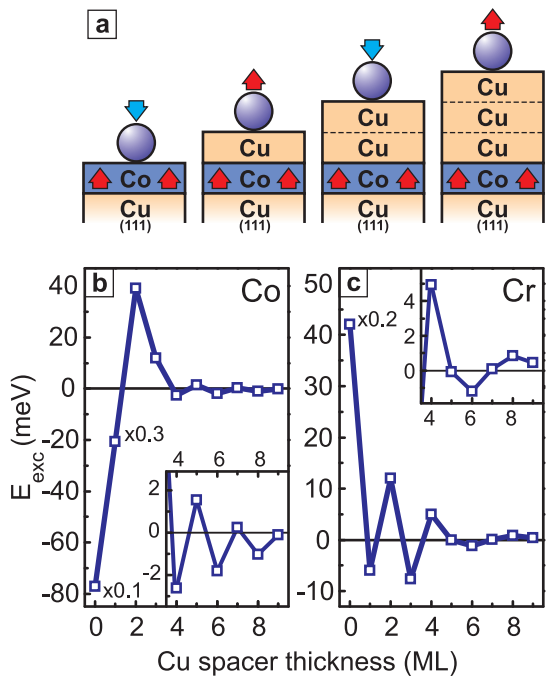}
        \caption{\label{fig:1}(color online) (a) The setup for calculations: an adatom coupled to a Co layer through a nonmagnetic Cu spacer of a varying thickness. Exchange coupling energies of a Co (b) and Cr (c) adatoms versus spacer thickness. First several points of the curves were scaled down for clarity. The scaling factors are given next to respective data points. The insets in each graph show the respective curves on a smaller scale at larger spacer thicknesses.}
	\end{figure}

    To study the interplay between the exchange interaction and the system's geometry we choose a Cu(111) surface as a base for our calculations. Without limiting the generality we consider, as the first and simplest model system, single magnetic 3d adatoms (Co and Cr) placed on top of a Cu spacers of various thicknesses covering a single Co monolayer (ML) (Fig.~\ref{fig:1}a). Our choice of geometry and atomic species is governed by the fact, that thin Co films are known to have out-of-plain magnetization as an inherent property \cite{CoMLGro1}. This fact provides for an increased surface symmetry for the orientation of adatom and addimer spins. The Co layer thickness of 1 monolayer (ML) was chosen as a marginal case and also should not affect the generality of results and conclusions.

    The dependence of the exchange interaction energy of an adatom on the thickness of the spacer is shown in Fig.~\ref{fig:1}b for Co and Fig.~\ref{fig:1}c for Cr. The presence of oscillations in the exchange coupling energies, similar to those observed for the interlayer exchange coupling \cite{Parkin1,BrunoIEC_PRL91}, can be attributed to the effect of quantum confinement in the overlayer. The presence of the vacuum barrier on one side of the overlayer and the magnetic layer on the other causes the electrons to become effectively confined between them \cite{BrunoIEC_PRB95}. Moreover, due to the ferromagnetic nature of the Co monolayer, the confinement of majority and minority electrons will be different, causing the formation of spin-polarized interference patterns. The magnetic properties of adatoms adsorbed on an overlayer are inevitably affected by the changes in the conduction electron densities.

    The coupling energies presented in Fig.~\ref{fig:1}(b and c) suggest that by coupling the spins of adatoms to that of a monolayer one gets a reliable means of stabilizing single atomic spins on the surface in either a ferro- or an antiferromagnetic configuration. The switching between configurations can be done by adjusting the thickness of the overlayer.

    \begin{figure}
		\includegraphics{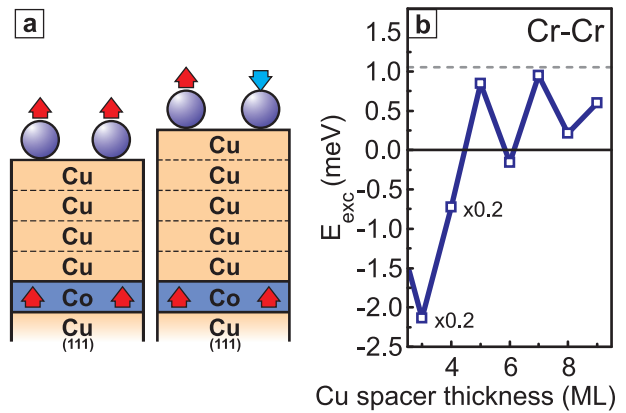}
        \caption{\label{fig:2}(color online). System under consideration: Cr adatoms coupled to a Co monolayer across a Cu spacer of a varying thickness~(a). Exchange coupling of a Cr adatoms at $7.66~\mathrm{\AA}$ separation aligned along the $[\overline{1}10]$ direction of a Cu(111) surface versus spacer thickness (b). Due to the energetic non-degeneracy of $\uparrow\uparrow$ and $\downarrow\downarrow$ configurations the exchange energies were calculated as follows $E_{exc}=min(E_{\uparrow\uparrow},E_{\downarrow\downarrow})-E_{\uparrow\downarrow}$. The gray dashed line gives the level of the exchange coupling strength between Cr atoms on a clean Cu(111) surface. First two points of the curve were scaled down for clarity. The scaling factors are given next to respective data points.}
	\end{figure}

    Let us now move forward, to the second example of spin-dependent scattering affecting the exchange interaction in the system. Up to now we have only considered single adatoms on the surface. If we now add a second adatom to the system, thus creating a dimer (Fig.~\ref{fig:2}a), we will find that the orientation of each of the atomic spins is determined by the competition between two exchange couplings: a coupling to the magnetic monolayer and the interatomic coupling in the dimer. Consequently, precise control over the thickness of the spacer provides us with an additional degree of freedom in adjusting the exchange interaction between single adatoms at any separation. At smaller spacer thicknesses where the coupling of a single adatom to the monolayer prevails the monolayer acts as a stabilizing element, rigidly fixing the dimer in either a $\uparrow\uparrow$ or a $\downarrow\downarrow$ configuration. At larger spacer thicknesses, when the interatomic exchange energy becomes comparable to that of the coupling to the monolayer, the system's spins become most susceptible to manipulations by changing the spacer thickness and interatomic separation. As an example the exchange interaction energy of a Cr-Cr dimer at $7.66~\mathrm{\AA}$ separation is shown in Fig.~\ref{fig:2}b as a function of the spacer thickness. It is clear that by adjusting the number of monolayers in the spacer one can tune the dimer to have an exchange coupling ranging from a strong ferromagnetic (at 1-4 ML) to an antiferromagnetic one ($5,7~\mathrm{ML}$).

    \begin{figure}
        \includegraphics{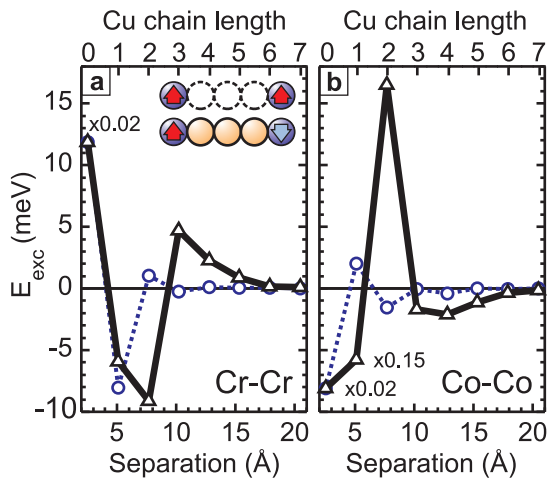}
        \caption{\label{fig:3}(color online). (a) The exchange interaction between two single Cr atoms on a Cu(111) surface (dotted blue line, open circle) and the exchange interaction between those atoms linked with a nonmagnetic Cu chain (solid black line, open triangles). (b) The same dependencies for a Co pair.
        The exchange energy is calculated via the following formula $E_{exc}=E_{\uparrow\uparrow}-E_{\uparrow\downarrow}$. First several points of the curves were rescaled for clarity. The scaling factors are given next to respective data points.}
    \end{figure}

    To demonstrate how artificial nonmagnetic nanostructures introduced in between a pair of magnetic impurities on a Cu(111) surface affect the spin alignment within the dimer we choose, as the third system of our studies, close-packed Cu chains linking a pair of magnetic impurities, following the work of Lagoute et al. \cite{Folsch07}. The total chain length in our calculation is varying from 2 to 15 atoms. Calculations for a pair of single adatoms were performed for the same positions of magnetic impurities as were used in the case of chains.

    Exchange interactions between two single Cr adatoms and in Cr pairs linked with Cu chains are plotted in Fig.~\ref{fig:3}a with the dotted blue line and the solid black line respectively.  Nearest neighbor Cr dimer exhibits strong antiferromagnetic coupling. The exchange interaction of a Cr pair at a $5.11~\mathrm{\AA}$ separation reverse its sign to ferromagnetic, but its magnitude is strongly reduced because a direct overlap of  wave functions of Cr adatoms is not possible. At further separations the exchange interaction oscillating decays to zero. The decay rate at large separations is determined by the two dimensional free-electron-like electron gas stemming from the surface state arising in the projected bulk band gap of Cu(111). It was demonstrated that the exchange interaction in such kind of systems oscillates at large distances with a period determined by the surface state momentum at the Fermi energy \cite{Stepanyuk04}. Amplitude of these oscillations decays proportional to $r^{-2}$, where $r$ is the adatom-adatom separation.

    Let us trace now the effect of a Cu chain. The only Cu atom inserted in between Cr dimer at $5.11~\mathrm{\AA}$ separation slightly reduces the exchange interaction in comparison to the reference case of single Cr adatoms. But longer mixed chains exhibit drastic differences from the reference case. For instance, two Cu atoms linking the Cr pair reverse the sign of the exchange interaction and strongly enhance its magnitude to $9~\mathrm{meV}$. The system with three Cu atoms in the chain has the antiferromagnetic exchange interaction equal to $5~\mathrm{meV}$ which is about 10 times larger than the magnitude of ferromagnetic exchange interaction of the reference system. Further elongation of the chain length results in rapid decay of the exchange interaction.

    Profound effect of short chains on the exchange interaction and its decay in long ones can be explained by competition of two factors. On the one hand, the density of conduction electrons at Cu atoms is much higher than that of the surface state, so interference effects are expected to be more pronounced and therefore exchange energies can reasonably be higher. On the other hand, each Cu atom scatters conduction electrons to the substrate, so interference patterns should rapidly decay with the increase of Cu chain length.

    Figure \ref{fig:3}b shows the effect of a Cu chain on the exchange interaction between Co adatoms. A close packed Co dimer on Cu(111) exhibits a strong ferromagnetic coupling with a strength of $0.4~\mathrm{eV}$. It is evident that short Cu chains significantly affect the exchange interaction as it has just been demonstrated for the case of Cr magnetic impurities. A single Cu atom inserted between Co atoms in a dimer at $5.11~\mathrm{\AA}$ separation changes the sign of the exchange interaction and enhances its magnitude. Two Cu atoms permit one to get an antiferromagnetically coupled Co pair. Further increase of the chain length reverses the sign of the exchange interaction back to ferromagnetic. Magnitude of the exchange interaction in presence of a linking chain is of the order higher than that for the reference case. Similar to the case of Cr impurities, magnitude of the exchange interaction between Co coupled through the Cu chain rapidly decay in large chains.

    \begin{figure}
        \includegraphics{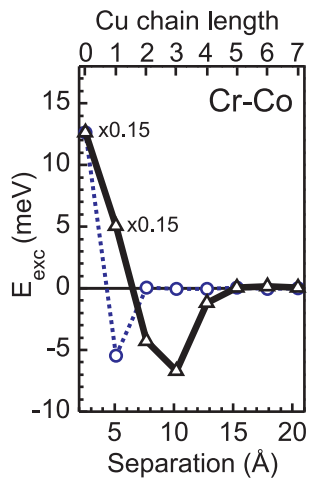}
        \caption{\label{fig:4}(color online). The exchange interaction between Co and Cr adatoms on a Cu(111) surface (dotted blue line, open circle) and the exchange interaction between those atoms linked with nonmagnetic Cu chains (solid black line, open triangles). First several points of the curve were scaled down for clarity. The scaling factors are given next to respective data points.}
    \end{figure}

    Yet another way to tailor the exchange interaction is to use mixed pairs of magnetic impurities. In Fig.~\ref{fig:4} we present our results for Cr-Co pairs. Co and Cr in a close packed dimer are coupled antiferromagnetically with an energy of $78~\mathrm{meV}$. This energy is almost one order of magnitude lower than that for a homonuclear Cr dimer. Non-linked Co and Cr adatoms at $5.11~\mathrm{\AA}$ are coupled ferromagnetically with an energy of $5.5~\mathrm{meV}$. At larger distances their exchange interaction is negligibly small. Short Cu chains inserted between Co and Cr significantly enhance the magnitude of the exchange interaction. A single Cu atom placed between Co and Cr reduces the antiferromagnetic exchange interaction to $30~\mathrm{meV}$. It is important to note that the same structures with pure Co and Cr dimers exhibit exchange interactions of the same order but of the opposite sign. The next points of the graph also demonstrate significant differences from cases of homogeneous magnetic pairs. Magnetic impurities of different species thus provides a wide range possibilities in tailoring the exchange interaction.


    In conclusion, we have shown that the exchange coupling of an adatom to a monolayer across a paramagnetic spacer oscillates with the thickness of the latter. This provides reliable means for stabilizing the spin of the adatom in either a ferromagnetic or an antiferromagnetic configuration with respect to the magnetic moment of the monolayer. The competition between interatomic and atom-layer couplings in a dimer allows one, by adjusting the overlayer thickness and the interatomic separation, to tailor the exchange coupling between single magnetic impurities on a surface. Moreover, further modification of the system's magnetic properties can be achieved by introducing nonmagnetic atomic chains in between the adsorbed adatoms, thus changing the propagation medium for the conduction electrons, that mediate the exchange coupling in the system. Atomic chains can enhance the exchange interaction between magnetic impurities, can change the sign of the exchange interaction or quench in completely.

    The presented approach provides a theoretically transparent and technologically feasible way of tailoring single spins in magnetic nanostructures adsorbed on a surface, which can prove invaluable for spintronic applications.

    This work was supported by Deutsche Forschungsgemeinschaft (DFG SPP 1165 and SSP 1153).


\begin{thebibliography}{22}
    \expandafter\ifx\csname natexlab\endcsname\relax\def\natexlab#1{#1}\fi
    \expandafter\ifx\csname bibnamefont\endcsname\relax
      \def\bibnamefont#1{#1}\fi
    \expandafter\ifx\csname bibfnamefont\endcsname\relax
      \def\bibfnamefont#1{#1}\fi
    \expandafter\ifx\csname citenamefont\endcsname\relax
      \def\citenamefont#1{#1}\fi
    \expandafter\ifx\csname url\endcsname\relax
      \def\url#1{\texttt{#1}}\fi
    \expandafter\ifx\csname urlprefix\endcsname\relax\def\urlprefix{URL }\fi
    \providecommand{\bibinfo}[2]{#2}
    \providecommand{\eprint}[2][]{\url{#2}}

    \bibitem[{\citenamefont{Loss and DiVincenzo}(1998)}]{QuantComp}
    \bibinfo{author}{\bibfnamefont{D.}~\bibnamefont{Loss}} \bibnamefont{and}
      \bibinfo{author}{\bibfnamefont{D.~P.} \bibnamefont{DiVincenzo}},
      \bibinfo{journal}{Phys. Rev. A} \textbf{\bibinfo{volume}{57}},
      \bibinfo{pages}{120} (\bibinfo{year}{1998}).

    \bibitem[{\citenamefont{Binasch et~al.}(1989)\citenamefont{Binasch, Gr\"unberg,
      Saurenbach, and Zinn}}]{Grunberg2}
    \bibinfo{author}{\bibfnamefont{G.}~\bibnamefont{Binasch}},
      \bibinfo{author}{\bibfnamefont{P.}~\bibnamefont{Gr\"unberg}},
      \bibinfo{author}{\bibfnamefont{F.}~\bibnamefont{Saurenbach}},
      \bibnamefont{and} \bibinfo{author}{\bibfnamefont{W.}~\bibnamefont{Zinn}},
      \bibinfo{journal}{Phys. Rev. B} \textbf{\bibinfo{volume}{39}},
      \bibinfo{pages}{4828} (\bibinfo{year}{1989}).

    \bibitem[{\citenamefont{Baibich et~al.}(1988)\citenamefont{Baibich, Broto,
      Fert, Van~Dau, Petroff, Eitenne, Creuzet, Friederich, and Chazelas}}]{Fert}
    \bibinfo{author}{\bibfnamefont{M.~N.} \bibnamefont{Baibich}},
      \bibinfo{author}{\bibfnamefont{J.~M.} \bibnamefont{Broto}},
      \bibinfo{author}{\bibfnamefont{A.}~\bibnamefont{Fert}},
      \bibinfo{author}{\bibfnamefont{F.~N.} \bibnamefont{Van~Dau}},
      \bibinfo{author}{\bibfnamefont{F.}~\bibnamefont{Petroff}},
      \bibinfo{author}{\bibfnamefont{P.}~\bibnamefont{Eitenne}},
      \bibinfo{author}{\bibfnamefont{G.}~\bibnamefont{Creuzet}},
      \bibinfo{author}{\bibfnamefont{A.}~\bibnamefont{Friederich}},
      \bibnamefont{and} \bibinfo{author}{\bibfnamefont{J.}~\bibnamefont{Chazelas}},
      \bibinfo{journal}{Phys. Rev. Lett.} \textbf{\bibinfo{volume}{61}},
      \bibinfo{pages}{2472} (\bibinfo{year}{1988}).

    \bibitem[{\citenamefont{Barna\ifmmode~\acute{s}\else \'{s}\fi{} and
      Bruynseraede}(1996)}]{PhysRevB.53.R2956}
    \bibinfo{author}{\bibfnamefont{J.}~\bibnamefont{Barna\ifmmode~\acute{s}\else
      \'{s}\fi{}}} \bibnamefont{and}
      \bibinfo{author}{\bibfnamefont{Y.}~\bibnamefont{Bruynseraede}},
      \bibinfo{journal}{Phys. Rev. B} \textbf{\bibinfo{volume}{53}},
      \bibinfo{pages}{R2956} (\bibinfo{year}{1996}).

    \bibitem[{\citenamefont{Bruno and Chappert}(1991)}]{BrunoIEC_PRL91}
    \bibinfo{author}{\bibfnamefont{P.}~\bibnamefont{Bruno}} \bibnamefont{and}
      \bibinfo{author}{\bibfnamefont{C.}~\bibnamefont{Chappert}},
      \bibinfo{journal}{Phys. Rev. Lett.} \textbf{\bibinfo{volume}{67}},
      \bibinfo{pages}{1602} (\bibinfo{year}{1991}).

    \bibitem[{\citenamefont{Bruno and Chappert}(1992)}]{BrunoIEC_PRB92}
    \bibinfo{author}{\bibfnamefont{P.}~\bibnamefont{Bruno}} \bibnamefont{and}
      \bibinfo{author}{\bibfnamefont{C.}~\bibnamefont{Chappert}},
      \bibinfo{journal}{Phys. Rev. B} \textbf{\bibinfo{volume}{46}},
      \bibinfo{pages}{261} (\bibinfo{year}{1992}).

    \bibitem[{\citenamefont{Stroscio et~al.}(2006)\citenamefont{Stroscio, Tavazza,
      Crain, Celotta, and Chaka}}]{Stroscio06}
    \bibinfo{author}{\bibfnamefont{J.~A.} \bibnamefont{Stroscio}},
      \bibinfo{author}{\bibfnamefont{F.}~\bibnamefont{Tavazza}},
      \bibinfo{author}{\bibfnamefont{J.~N.} \bibnamefont{Crain}},
      \bibinfo{author}{\bibfnamefont{R.~J.} \bibnamefont{Celotta}},
      \bibnamefont{and} \bibinfo{author}{\bibfnamefont{A.~M.} \bibnamefont{Chaka}},
      \bibinfo{journal}{Science} \textbf{\bibinfo{volume}{313}},
      \bibinfo{pages}{948} (\bibinfo{year}{2006}).

    \bibitem[{\citenamefont{Lagoute et~al.}(2007)\citenamefont{Lagoute, Nacci, and
      F\"olsch}}]{Folsch07}
    \bibinfo{author}{\bibfnamefont{J.}~\bibnamefont{Lagoute}},
      \bibinfo{author}{\bibfnamefont{C.}~\bibnamefont{Nacci}}, \bibnamefont{and}
      \bibinfo{author}{\bibfnamefont{S.}~\bibnamefont{F\"olsch}},
      \bibinfo{journal}{Phys. Rev. Lett.} \textbf{\bibinfo{volume}{98}},
      \bibinfo{eid}{146804} (\bibinfo{year}{2007}).

    \bibitem[{\citenamefont{Hirjibehedin et~al.}(2006)\citenamefont{Hirjibehedin,
      Lutz, and Heinrich}}]{Hirjibehedin:Nature312}
    \bibinfo{author}{\bibfnamefont{C.~F.} \bibnamefont{Hirjibehedin}},
      \bibinfo{author}{\bibfnamefont{C.~P.} \bibnamefont{Lutz}}, \bibnamefont{and}
      \bibinfo{author}{\bibfnamefont{A.~J.} \bibnamefont{Heinrich}},
      \bibinfo{journal}{Science} \textbf{\bibinfo{volume}{312}},
      \bibinfo{pages}{1021} (\bibinfo{year}{2006}).

    \bibitem[{\citenamefont{Wahl et~al.}(2007)\citenamefont{Wahl, Simon,
      Diekh\"{o}ner, Stepanyuk, Bruno, Schneider, and Kern}}]{Kern07}
    \bibinfo{author}{\bibfnamefont{P.}~\bibnamefont{Wahl}},
      \bibinfo{author}{\bibfnamefont{P.}~\bibnamefont{Simon}},
      \bibinfo{author}{\bibfnamefont{L.}~\bibnamefont{Diekh\"{o}ner}},
      \bibinfo{author}{\bibfnamefont{V.~S.} \bibnamefont{Stepanyuk}},
      \bibinfo{author}{\bibfnamefont{P.}~\bibnamefont{Bruno}},
      \bibinfo{author}{\bibfnamefont{M.~A.} \bibnamefont{Schneider}},
      \bibnamefont{and} \bibinfo{author}{\bibfnamefont{K.}~\bibnamefont{Kern}},
      \bibinfo{journal}{Phys. Rev. Lett.} \textbf{\bibinfo{volume}{98}},
      \bibinfo{eid}{056601} (\bibinfo{year}{2007}).

    \bibitem[{\citenamefont{Meier et~al.}(2008)\citenamefont{Meier, Zhou, Wiebe,
      and Wiesendanger}}]{FockoMeier04042008}
    \bibinfo{author}{\bibfnamefont{F.}~\bibnamefont{Meier}},
      \bibinfo{author}{\bibfnamefont{L.}~\bibnamefont{Zhou}},
      \bibinfo{author}{\bibfnamefont{J.}~\bibnamefont{Wiebe}}, \bibnamefont{and}
      \bibinfo{author}{\bibfnamefont{R.}~\bibnamefont{Wiesendanger}},
      \bibinfo{journal}{Science} \textbf{\bibinfo{volume}{320}},
      \bibinfo{pages}{82} (\bibinfo{year}{2008}).

    \bibitem[{\citenamefont{Wildberger et~al.}(1995)\citenamefont{Wildberger,
      Stepanyuk, Lang, Zeller, and Dederichs}}]{Stepanyuk.PRL.1995}
    \bibinfo{author}{\bibfnamefont{K.}~\bibnamefont{Wildberger}},
      \bibinfo{author}{\bibfnamefont{V.~S.} \bibnamefont{Stepanyuk}},
      \bibinfo{author}{\bibfnamefont{P.}~\bibnamefont{Lang}},
      \bibinfo{author}{\bibfnamefont{R.}~\bibnamefont{Zeller}}, \bibnamefont{and}
      \bibinfo{author}{\bibfnamefont{P.~H.} \bibnamefont{Dederichs}},
      \bibinfo{journal}{Phys. Rev. Lett.} \textbf{\bibinfo{volume}{75}},
      \bibinfo{pages}{509} (\bibinfo{year}{1995}).

    \bibitem[{\citenamefont{Zeller et~al.}(1995)\citenamefont{Zeller, Dederichs,
      \'Ujfalussy, Szunyogh, and Weinberger}}]{Zeller.PRB.1995}
    \bibinfo{author}{\bibfnamefont{R.}~\bibnamefont{Zeller}},
      \bibinfo{author}{\bibfnamefont{P.~H.} \bibnamefont{Dederichs}},
      \bibinfo{author}{\bibfnamefont{B.}~\bibnamefont{\'Ujfalussy}},
      \bibinfo{author}{\bibfnamefont{L.}~\bibnamefont{Szunyogh}}, \bibnamefont{and}
      \bibinfo{author}{\bibfnamefont{P.}~\bibnamefont{Weinberger}},
      \bibinfo{journal}{Phys. Rev. B} \textbf{\bibinfo{volume}{52}},
      \bibinfo{pages}{8807} (\bibinfo{year}{1995}).

    \bibitem[{\citenamefont{Zabloudil et~al.}(2005)\citenamefont{Zabloudil,
      Hammerling, Szunyogh, and Weinberger}}]{zab05}
    \bibinfo{author}{\bibfnamefont{J.}~\bibnamefont{Zabloudil}},
      \bibinfo{author}{\bibfnamefont{R.}~\bibnamefont{Hammerling}},
      \bibinfo{author}{\bibfnamefont{L.}~\bibnamefont{Szunyogh}}, \bibnamefont{and}
      \bibinfo{author}{\bibfnamefont{P.}~\bibnamefont{Weinberger}},
      \bibinfo{journal}{\textit{Electron Scattering in Solid Matter}, Springer
      Series in Solid-State Sciences Vol. 147}  (\bibinfo{year}{2005}).

    \bibitem[{\citenamefont{Lang et~al.}(1996)\citenamefont{Lang, Nordstr\"om,
      Wildberger, Zeller, Dederichs, and Hoshino}}]{Lang96}
    \bibinfo{author}{\bibfnamefont{P.}~\bibnamefont{Lang}},
      \bibinfo{author}{\bibfnamefont{L.}~\bibnamefont{Nordstr\"om}},
      \bibinfo{author}{\bibfnamefont{K.}~\bibnamefont{Wildberger}},
      \bibinfo{author}{\bibfnamefont{R.}~\bibnamefont{Zeller}},
      \bibinfo{author}{\bibfnamefont{P.~H.} \bibnamefont{Dederichs}},
      \bibnamefont{and} \bibinfo{author}{\bibfnamefont{T.}~\bibnamefont{Hoshino}},
      \bibinfo{journal}{Phys. Rev. B} \textbf{\bibinfo{volume}{53}},
      \bibinfo{pages}{9092} (\bibinfo{year}{1996}).

    \bibitem[{\citenamefont{Stepanyuk et~al.}(2004)\citenamefont{Stepanyuk,
      Niebergall, Longo, Hergert, and Bruno}}]{Stepanyuk04}
    \bibinfo{author}{\bibfnamefont{V.~S.} \bibnamefont{Stepanyuk}},
      \bibinfo{author}{\bibfnamefont{L.}~\bibnamefont{Niebergall}},
      \bibinfo{author}{\bibfnamefont{R.~C.} \bibnamefont{Longo}},
      \bibinfo{author}{\bibfnamefont{W.}~\bibnamefont{Hergert}}, \bibnamefont{and}
      \bibinfo{author}{\bibfnamefont{P.}~\bibnamefont{Bruno}},
      \bibinfo{journal}{Phys. Rev. B} \textbf{\bibinfo{volume}{70}},
      \bibinfo{pages}{075414} (\bibinfo{year}{2004}).

    \bibitem[{\citenamefont{Hyldgaard and Persson}(2000)}]{Hylgaard00}
    \bibinfo{author}{\bibfnamefont{P.}~\bibnamefont{Hyldgaard}} \bibnamefont{and}
      \bibinfo{author}{\bibfnamefont{M.}~\bibnamefont{Persson}},
      \bibinfo{journal}{J. Phys. Cond. Matt.} \textbf{\bibinfo{volume}{12}},
      \bibinfo{pages}{L13} (\bibinfo{year}{2000}).

    \bibitem[{\citenamefont{Stepanyuk et~al.}(2005)\citenamefont{Stepanyuk,
      Niebergall, Hergert, and Bruno}}]{Stepanyuk05}
    \bibinfo{author}{\bibfnamefont{V.~S.} \bibnamefont{Stepanyuk}},
      \bibinfo{author}{\bibfnamefont{L.}~\bibnamefont{Niebergall}},
      \bibinfo{author}{\bibfnamefont{W.}~\bibnamefont{Hergert}}, \bibnamefont{and}
      \bibinfo{author}{\bibfnamefont{P.}~\bibnamefont{Bruno}},
      \bibinfo{journal}{Phys. Rev. Lett.} \textbf{\bibinfo{volume}{94}},
      \bibinfo{eid}{187201} (\bibinfo{year}{2005}).

    \bibitem[{\citenamefont{Zheng et~al.}(2000)\citenamefont{Zheng, Shen, Barthel,
      Ohresser, Mohan, and Kirschner}}]{CoMLGro1}
    \bibinfo{author}{\bibfnamefont{M.}~\bibnamefont{Zheng}},
      \bibinfo{author}{\bibfnamefont{J.}~\bibnamefont{Shen}},
      \bibinfo{author}{\bibfnamefont{J.}~\bibnamefont{Barthel}},
      \bibinfo{author}{\bibfnamefont{P.}~\bibnamefont{Ohresser}},
      \bibinfo{author}{\bibfnamefont{C.~V.} \bibnamefont{Mohan}}, \bibnamefont{and}
      \bibinfo{author}{\bibfnamefont{J.}~\bibnamefont{Kirschner}},
      \bibinfo{journal}{J. Phys.: Cond. Matt.} \textbf{\bibinfo{volume}{12}},
      \bibinfo{pages}{783} (\bibinfo{year}{2000}).

    \bibitem[{\citenamefont{Parkin et~al.}(1990)\citenamefont{Parkin, More, and
      Roche}}]{Parkin1}
    \bibinfo{author}{\bibfnamefont{S.~S.~P.} \bibnamefont{Parkin}},
      \bibinfo{author}{\bibfnamefont{N.}~\bibnamefont{More}}, \bibnamefont{and}
      \bibinfo{author}{\bibfnamefont{K.~P.} \bibnamefont{Roche}},
      \bibinfo{journal}{Phys. Rev. Lett.} \textbf{\bibinfo{volume}{64}},
      \bibinfo{pages}{2304} (\bibinfo{year}{1990}).

    \bibitem[{\citenamefont{Bruno}(1995)}]{BrunoIEC_PRB95}
    \bibinfo{author}{\bibfnamefont{P.}~\bibnamefont{Bruno}},
      \bibinfo{journal}{Phys. Rev. B} \textbf{\bibinfo{volume}{52}},
      \bibinfo{pages}{411} (\bibinfo{year}{1995}).
\end{thebibliography}
\end{document}